\documentstyle[12pt]{article}

\def\hybrid{\topmargin -20pt    \oddsidemargin 0pt  
        \headheight 0pt \headsep 0pt  
        \textwidth 6.25in       
        \textheight 9.5in       
        \marginparwidth .875in  
        \parskip 5pt plus 1pt   \jot = 1.5ex}  

\def\ket#1{|{#1}\rangle}  
\def\noi{\noindent}  
  
\def\baselinestretch{1.2}  
  
\catcode`\@=11  
  
\def\marginnote#1{}  
\def\draftlabel#1{{\@bsphack\if@filesw {\let\thepage\relax  
   \xdef\@gtempa{\write\@auxout{\string  
      \newlabel{#1}{{\@currentlabel}{\thepage}}}}}\@gtempa  
   \if@nobreak \ifvmode\nobreak\fi\fi\fi\@esphack}  
        \gdef\@eqnlabel{#1}}  
\def\@eqnlabel{}  
\def\@vacuum{}  
\def\draftmarginnote#1{\marginpar{\raggedright\scriptsize\tt#1}}  
  
\def\draft{\oddsidemargin -.2truein  
        \def\@oddfoot{\sl preliminary draft \hfil  
        \rm\thepage\hfil\sl\today\quad\militarytime}  
        \let\@evenfoot\@oddfoot \overfullrule 3pt  
        \let\label=\draftlabel  
        \let\marginnote=\draftmarginnote  
   \def\@eqnnum{(\theequation)\rlap{\kern\marginparsep\tt\@eqnlabel}%
\global\let\@eqnlabel\@vacuum}  }  
  
\def\preprint{\twocolumn\sloppy\flushbottom\parindent 2em  
        \leftmargini 2em\leftmarginv .5em\leftmarginvi .5em  
        \oddsidemargin -.5in    \evensidemargin -.5in  
        \columnsep .4in \footheight 0pt  
        \textwidth 10.in        \topmargin  -.4in  
        \headheight 12pt \topskip .4in  
        \textheight 6.9in \footskip 0pt  
        \def\@oddhead{\thepage\hfil\addtocounter{page}{1}\thepage}  
        \let\@evenhead\@oddhead \def\@oddfoot{} \def\@evenfoot{} }  
  
\def\numberbysection{\@addtoreset{equation}{section}  
        \def\theequation{\thesection.\arabic{equation}}}  
  
\def\underline#1{\relax\ifmmode\@@underline#1\else  
        $\@@underline{\hbox{#1}}$\relax\fi}

\def\titlepage{\@restonecolfalse\if@twocolumn\@restonecoltrue  
\onecolumn  
     \else \newpage \fi \thispagestyle{empty}\c@page\z@  
        \def\thefootnote{\fnsymbol{footnote}} }  
  
\def\endtitlepage{\if@restonecol\twocolumn \else \newpage \fi  
        \def\thefootnote{\arabic{footnote}}  
        \setcounter{footnote}{0}}  
  
\catcode`@=12  
\relax  
  
\def\figcap{\section*{Figure Captions\markboth  
        {FIGURECAPTIONS}{FIGURECAPTIONS}}\list  
        {Figure \arabic{enumi}:\hfill}{\settowidth\labelwidth{Figure  
999:}  
        \leftmargin\labelwidth  
        \advance\leftmargin\labelsep\usecounter{enumi}}}  
 \relax  
\def\tablecap{\section*{Table Captions\markboth  
        {TABLECAPTIONS}{TABLECAPTIONS}}\list  
        {Table \arabic{enumi}:\hfill}{\settowidth\labelwidth{Table  
999:}  
        \leftmargin\labelwidth  
        \advance\leftmargin\labelsep\usecounter{enumi}}}  
 \relax  
\def\reflist{\section*{References\markboth  
        {REFLIST}{REFLIST}}\list  
        {[\arabic{enumi}]\hfill}{\settowidth\labelwidth{[999]}  
        \leftmargin\labelwidth  
        \advance\leftmargin\labelsep\usecounter{enumi}}}  
 \relax  
\makeatletter  
\newcounter{pubctr}  
\def\publist{\@ifnextchar[{\@publist}{\@@publist}}  
\def\@publist[#1]{\list  
        {[\arabic{pubctr}]\hfill}{\settowidth\labelwidth{[999]}  
        \leftmargin\labelwidth  
        \advance\leftmargin\labelsep  
        \@nmbrlisttrue\def\@listctr{pubctr}  
        \setcounter{pubctr}{#1}\addtocounter{pubctr}{-1}}}  
\def\@@publist{\list  
        {[\arabic{pubctr}]\hfill}{\settowidth\labelwidth{[999]}  
        \leftmargin\labelwidth  
        \advance\leftmargin\labelsep  
        \@nmbrlisttrue\def\@listctr{pubctr}}}  
 \relax  
\makeatother  
\newskip\humongous \humongous=0pt plus 1000pt minus 1000pt

\newif\ifdtup

\font\Scbig=cmss10 scaled\magstep1  
\font\Scscr=cmss8 scaled\magstep1  
\font\Scscrscr=cmss8  
\newfam\Scfam  
\textfont\Scfam=\Scbig  
\scriptfont\Scfam=\Scscr  
\scriptscriptfont\Scfam=\Scscrscr

\relax  
\hybrid  
\def\lvm{\leavevmode\hbox to\parindent{\hfill}}  
  
\def\thefootnote{\fnsymbol{footnote}}  
\def\BE{\begin{equation}}  
\def\EE{\end{equation}}  
\def\BA{\begin{eqnarray}}  
\def\EA{\end{eqnarray}}

\def\tt{\bar\tau}  
  
\def\lvm{\leavevmode\hbox to\parindent{\hfill}}  
\def\bar{\overline}

\def\BE{\begin{equation}}  
\def\EE{\end{equation} \vskip 0.30\baselineskip}  
\def\BA{\begin{array}}  
\def\EA{\end{array}}  
  
\def\noi{\noindent}  
  
\def\frac#1#2{{\textstyle{{#1}\over{#2}}}}

\def\ket#1{|{#1}\rangle}

\newif\ifold \oldtrue   
\let\ssection=\section  
\def\section{\setcounter{equation}{0}\ssection}  
  
\begin{document}  
\renewcommand{\theequation}{\arabic{equation}}  
\newcommand{\beq}{\begin{equation}}  
\newcommand{\eeq}[1]{\label{#1}\end{equation}}  
\newcommand{\ber}{\begin{eqnarray}}  
\newcommand{\eer}[1]{\label{#1}\end{eqnarray}}  
\begin{titlepage}  
\begin{center}  
   
\hfill IMAFF-FM-06-23\\
\hfill NIKHEF/2006-006\\
\hfill hep-th/0612305
\vskip 1in
  
{\large \bf Remarks on Global Anomalies in DHS Orientifolds}

\vskip 1in 
 
{\bf Beatriz Gato-Rivera}\footnote{Also known as B. Gato}\\
\vskip .5in

{\em Instituto de Matem\'aticas y F\'\i sica Fundamental, CSIC, \\
Serrano 123, Madrid 28006, Spain}\\

{\em NIKHEF Theory Group, Kruislaan 409, \\
1098 SJ Amsterdam, The Netherlands}\\

\vskip 1in

\end{center}
  
\begin{center} {\bf ABSTRACT } \end{center}  
\begin{quotation}  
I review the main features of the Dijkstra--Huiszoon--Schellekens (DHS) 
orientifolds and report on the search for global anomalies that 
Schellekens and the author have performed for these models.
   
\end{quotation} 
 
\vskip 2cm  
  
December 2006  

\vskip 2cm

\noi 
Invited talk given at the `Strings 2006 Workshop', Shanghai (China), June 2006.

\end{titlepage}  
  
\def\baselinestretch{1.2}  
\baselineskip 17 pt

\section{Introduction}\lvm  
  
In this talk I will present some work done in collaboration with Bert Schellekens,
published in ref. \cite{Global}. In the first part of the talk I will introduce the
Dijkstra--Huiszoon--Schellekens (DHS) open strings. I will explain what are they,
how were they constructed, and their main features. In the second part I will
report on the search for global anomalies that we have performed for these models.
I will review briefly the subject of global (gauge) anomalies, then I will introduce 
the probe brane method to look for global anomalies in string/brane configurations
and finally I will present the results obtained for the DHS open strings. 

\section{Dijkstra--Huiszoon--Schellekens Open Strings}\lvm

The DHS open strings are a large number of supersymmetric open strings with
a chiral spectrum that exactly matches the supersymmetric Standard Model 
spectrum. They were constructed algebraically by the dutch group T. Dijkstra, 
L. Huiszoon and A.N. Schellekens \cite{DHS1} \cite{DHS2}, using the
technology of Rational Conformal Field Theory (RCFT) on surfaces with 
boundaries and crosscaps. To be precise, the DHS open strings were obtained 
building orientifolds based on tensor products of 
superconformal N=2 minimal models (Gepner models).

Let us say a few more words about these important subjects: RCFT on surfaces
with boundaries and crosscaps, orientifold constructions and Gepner models. 

\subsection{Some features}

\vskip 0.1in

CFT on surfaces with boundaries and crosscaps is the CFT description 
of the perturbative expansion of open string theories, the crosscaps being 
only necessary when the theory involves unoriented strings. The main 
ingredients in these theories are the one-loop worldsheets, which are the
2-dimensional surfaces swept out by the strings when moving in space-time. One 
finds four different types of topologically inequivalent one-loop worldsheets:
torus, Klein bottle, annulus, and M\"obius strip. Depending on the string theory 
at hand some of these surfaces may or may not appear, but for open unoriented 
strings the four types appear. These surfaces, called `direct channel' surfaces 
contain the information
about the complete spectrum of the CFT defined on them; that is, the
amplitudes of these surfaces represent the partition functions of the
theory that give the number of states level by level. But one can also
look at these surfaces from the `transverse channel' point of view,
that is exchanging the space and time coordinates of the worldsheet.
The torus does not change, but the annulus, the M\"obius strip and
the Klein bottle look quite different. The annulus is converted into a
cylinder between two boundaries, that shows the propagation of
closed strings between the two boundaries, the M\"obius strip is
converted into a cylinder between a boundary and a crosscap
(the M\"obius strip has only one boundary), and the Klein bottle turns into a 
cylinder between two crosscaps (a crosscap is a boundary with the opposite 
sides identified). The boundary states and the crosscap states are linear 
combinations of the so-called Ishibashi states
\BE
\ket{B_a} = \sum_i B_{ai} \ket{i}_B\, , \qquad
\ket{C} = \sum_i \Gamma_{i} \ket{i}_C
\EE
\noi
where $B_{ai}$ and $\Gamma_i$ are called boundary and crosscap coefficients, 
respectively. The label $a$ indicates that a CFT can have different sets 
of boundaries, whereas it can have only one type of crosscap.

The boundary and crosscap coefficients are very important quantities
since they contain information about the spectrum of the string states as well 
as information about the D-branes and orientifold planes. These coefficients are
constrained by  integrality and positivity conditions in order that the state
multiplicities of the spectra make sense. They are also subject to sewing 
constraints, world-sheet conditions needed to guarantee the correct factorization
of all amplitudes, which are rather difficult to solve.

When the basic CFT building blocks are rational, implying a finite number of
primary fields, rational conformal anomaly, and rational conformal weights,
one uses the term RCFT to denotes these theories.
The boundary and crosscap coefficients used in the DHS construction are based
on generic simple current\footnote{A pedagogical introduction to the subject of 
simple currents in CFT can be found in ref. \cite{Simple}} modifications of the 
Cardy boundary coefficients \cite{Cardy}  and the Rome group crosscap 
coefficients  \cite{Rome}\cite{RomeP1}:
\BE
B_{ai} = {S_{ai} \over \sqrt{S_{0i}}}\,,
 \qquad \Gamma_{i} = {P_{0i} \over \sqrt{S_{0i}}}\,.
\EE
\noi
where $S$ is the modular matrix, $S: \tau \rightarrow -1/{\tau}$ and P is defined 
as $P = \sqrt{T} S T^2 S \sqrt{T}$, $T$ being the modular matrix,
$T: \tau \rightarrow \tau + 1$.

A crucial consistency condition in CFT with boundaries and crosscaps 
is tadpole cancellation, a space-time constraint necessary to avoid infrared 
divergences in the one-loop amplitudes. This is very easy to see in the 
transverse channel where, by factorization, the `cylinder' decomposes as 
the product of the propagator times the tadpoles located at the extremes
(boundary tadpoles and/or crosscap tadpoles). If the tadpoles are left uncancelled
and correspond to physical states in the projected closed string spectrum, then
the system is unstable. If they do not correspond to physical states, however,
their presence implies a fundamental inconsistency in the theory which
translates into chiral anomalies in local gauge or gravitational symmetries.
Therefore the tadpoles must be cancelled, what implies that the Chan-Paton 
factors of the boundary states must be adjusted to some specific values. 
But the Chan-Paton factors reflect the gauge group of the theory and therefore the
tadpole cancellation fixes the possible allowed gauge groups.

\vskip 0.1in

An orientifold theory is obtained from a closed string theory by performing a
projection that inverts the orientation of the world-sheet. This often leads to
the apparition of open unoriented strings, what is most convenient because in
order to cancel the tadpoles one usually must add open strings to the original
closed string theory. For the orientifold construction leading to open unoriented
strings (also called `open descendant' construction) one starts with a closed
string theory with the corresponding torus partition function --  a modular 
invariant partition function to be precise --  then one computes 
boundary and crosscaps coefficients and next one builds the annulus, Klein 
bottle and M\"obius partition functions (see the details in ref. \cite{orientifold}).

\vskip 0.1in

As is well known, RCFT can be applied to obtain 4-dimensional
superstrings compactified on Calabi-Yau spaces. For this purpose one needs 
superconformal N=2 symmetry for the fields living on the worldsheets.
The simplest building blocks are superconformal N=2 minimal models and one 
can construct conformal tensors of them \cite{Gepner}, where the resulting
conformal anomaly $c_T$ is the sum of the conformal anomalies of each 
component. The value of  $c_T$ must be 9 so that the conformal anomaly 
of the corresponding string
theory vanishes. As a result, taking into account that for N=2 minimal models
$c=3k/(k+2)$ with $k=1,2,3,....$, one finds a total of 168 solutions of possible
tensor products satisfying the requirement $c_T = 9$.

\vskip .2in

\subsection{First results}
\vskip .1in

The orientifold construction used by DHS leads to 270.058 models compatible
with the supersymmetric Standard Model, at first sight. In particular, the
chiral spectrum of these models exactly matches the supersymmetric Standard 
Model spectrum, as was mentioned before.  The gauge groups of these models
usually, but not always, have hidden sectors in addition to the Standard Model 
gauge group $SU(3)$ x $SU(2)$ x $U(1)$. As a consequence, some of these models
differ only in hidden sector details.

\newpage

\section{Search for Global Anomalies}\lvm

\subsection{Global anomalies}
\vskip .1in

In certain cases global anomalies can occur even though all tadpole cancellation
conditions are satisfied. These are anomalies in the global definition of the field
theory path integral (gauge global anomalies to be precise), discovered by
Witten in 1982 \cite{Witten}. Global
anomalies are present when there is an odd number of massless fermions in the
vector representation of a symplectic factor of the gauge group (the best known
case are doublets of $SU(2)$, which is isomorphic to $Sp(2)$). In general, odd-rank
tensors of $Sp(2n)$ are problematic but in brane models only vectors occur.

The problem can be traced back to uncancelled K-theory charges of D-branes.
The reason is that tadpole cancellation only guarantees the cancellation of the
cohomology charges of the D-branes, characterized by long range RR fields
coupling to these charges, but D-branes may carry additional $Z_2$-charges
without a corresponding long range field. If these $Z_2$-charges remain
uncancelled, this manifests itself in the form of global anomalies. Unfortunately,
a complete description of global anomalies in theories of unoriented open strings
is not available at present. The best we can do is to examine if the symptoms
of the problem are present.

\vskip .2in

\subsection{The probe brane method}
\vskip .1in

In any given model, the presence or absence of global anomalies can be 
determined by simply counting the number of massless fermions in simplectic
vector representations, at the field theory level. However, there can be additional 
K-theory constraints rendering the theory globally inconsistent at the string level. 
To get a handle on these extra-constraints Uranga proposed a more general
method \cite{Uranga}. The basic idea is the following: {\it If to any model a 
symplectic brane-antibrane pair is added, the massless spectrum resulting from 
the intersections with the initial branes must be free from any additional global
anomalies}. The probe brane method also applies to local anomalies but gives 
nothing new if the tadpoles cancellation is satisfied.

\vskip .2in 

\subsection{Results}
\vskip .1in

We examined the 270.058 DHS spectra compatible with the Standard Model.
The total number of symplectic factors was found to be 845.513. Without 
using probe branes we found that only 1.015 spectra and 2.075 symplectic 
factors had global anomalies. Using probe branes we had to redo the 
computations. We simply searched the corresponding MIPFs (modular
invariant partition functions) again with the probe brane condition imposed
as an additional constraint. For the vast majority of MIPFs we did not
encounter any global probe brane anomalies for any solutions, even though
the number of potential anomalies was very large. From the 333 MIPFs
with tadpole cancellation solutions found by DHS, we encountered global
anomalies in only 25, and only for very few solutions in each case. 

As an example, let us consider the tensor product of the N=2 minimal models 
given by  $(k_1, k_2, k_3, k_4) = (1, 6, 46, 46)$. Now let us take the MIPF
number 10 (in our conventions) with 19.644 solutions.
One finds that the total number of new (independent) mod 2 conditions 
that must be satisfied for global anomaly cancellation is 147. Therefore 
there is a potential reduction of valid models of $2^{-147}$. In reality, 
however, from the original 19.644 solutions for this MIPF we
found that only 59 had to be removed due to global probe brane anomalies.

\vskip .2in 

\section{Conclusions and Final Remarks}
 \vskip .1in

We have seen that global anomalies, although very important in theory,
are almost irrelevant in practice for the DHS models. They only occur in
25 out of the 333 MIPFs with solutions, and for very few of these solutions. 
We must point out that the precise relation of these anomalies with 
K-theory charges remains to be understood.  Finally, it is an open question
if  the probe brane method is sufficient to trace all the global anomalies 
or whether there may be some that cannot be detected using this method.

\vskip 1cm   

\noi
{\bf Acknowledgements}

I thank the organizers of the Strings 2006 Shanghai Workshop for the
invitation to present some of my work. I also thank Bert Schellekens 
for many illuminating discussions about the subject of this talk. The
work presented has been partially supported by funding of the spanish
Ministerio de Educaci\' on y Ciencia, Research Project FPA2005-05046.

\vskip .2in 
\noi


\vskip .4in  
   
\begin{thebibliography}{9}  
\def\NPB{Nucl. Phys. B}  
\def\PLB{Phys. Lett. B}  
\def\MPLA{Mod. Phys. Lett. A}  
\def\IJMP{Int. J. Mod. Phys. }  

\bibitem{Global} B. Gato-Rivera and A.N. Schellekens, \PLB632 (2006) 728. 

\bibitem{DHS1} T.P.T. Dijkstra, L.R. Huiszoon and A.N. Schellekens, \PLB609 
(2005) 408.

\bibitem{DHS2} T.P.T. Dijkstra, L.R. Huiszoon and A.N. Schellekens, \NPB710
(2005) 3.

\bibitem{Simple} A.N. Schellekens and S. Yankielowicz, \IJMP 45 (1990) 2903.

\bibitem{Cardy} J. Cardy, \NPB324 (1989) 581.

\bibitem{Rome} A. Sagnotti and Y. S. Stanev, Fortsch. Phys. 44 (1996)
585 [Nucl. Phys. Proc. Suppl. 55B (1996) 200].

\bibitem{RomeP1} G. Pradisi, A. Sagnotti and Y. S. Stanev, \PLB354 
(1995) 279; \PLB356 (1995) 230.

\bibitem{orientifold} C. Angelantonj and A. Sagnotti, Phys. Rept. 371 (2002) 1,
Erratum-ibid 376 (2003) 339.

 \bibitem{Gepner} D. Gepner, \NPB296 (1988) 757.

\bibitem{Witten} E. Witten \PLB117 (1982) 324.

\bibitem{Uranga} A.M. Uranga, \NPB598 (2001) 225.

    
\end{thebibliography}
\end{document}